\documentclass[twocolumn,showpacs,preprintnumbers,amsmath,amssymb]{revtex4}
\usepackage{graphicx}% Include figure files
\usepackage{dcolumn}% Align table columns on decimal point
\usepackage{bm}% bold math

\begin{document}

\preprint{}

\title{Transverse self-interactions within an electron bunch moving in an arc of a circle (generalized)}
% Force line breaks with \\

\author{Gianluca Geloni}
\email{g.a.geloni@tue.nl}

\author{Jan Botman}
\author{Marnix van der Wiel}

\affiliation{%
Department of Applied Physics, Technische Universiteit Eindhoven,
\\
 P.O. Box 513, 5600MB Eindhoven, The Netherlands
}%

\author{Martin Dohlus}
\author{Evgeni Saldin}
\author{Evgeni Schneidmiller}
\affiliation{Deutsches Elektronen-Synchrotron DESY, \\
Notkestrasse 85, 22607 Hamburg, Germany
}%

\author{Mikhail Yurkov}
\affiliation{Particle Physics Laboratory (LSVE), Joint Institute for Nuclear Research, \\
141980 Dubna, Moscow Region, Russia}

\begin{abstract}
Transverse self-interactions within a "line" bunch of electrons
moving in an arc of a circle have been studied extensively by us
from an electrodynamical viewpoint. Here we treat an electron
bunch with a given vertical (i.e. perpendicular to the orbital
plane) size by a generalization of our previous work to the case
of a test particle with vertical displacement interacting with a
line bunch. In fact, since a bunch with vertical extent can always
be thought of as a superposition of displaced charge lines, all
the relevant physical aspects of the problem are included in the
study of that simple model. Our generalization results in a
physically meaningful and quantitative explanation of the
simulations obtained with the code $TraFiC^4$ as well as in
successful cross-checking of the code.
\end{abstract}

\pacs{29.27.Bd, 41.60.-m, 41.75.Ht}% PACS, the Physics and Astronomy

\begin{widetext}
\thispagestyle{empty}
\begin{large}
\textbf{DEUTSCHES ELEKTRONEN-SYNCHROTRON}\\
\end{large}

DESY 03-044

April 2003

\begin{eqnarray}
\nonumber &&\cr \nonumber && \cr \nonumber &&\cr
\end{eqnarray}

\begin{center}
\begin{Large}
\textbf{Transverse self-fields within an electron bunch moving
\\in an arc of a circle (generalized)}
\end{Large}
\begin{eqnarray}
\nonumber &&\cr \nonumber && \cr
\end{eqnarray}

\begin{large}
Gianluca Geloni, Jan Botman and Marnix van der Wiel
\end{large}

\textsl{\\Department of Applied Physics, Technische Universiteit
Eindhoven, \\P.O. Box 513, 5600MB Eindhoven, The Netherlands}
\begin{eqnarray}
\nonumber
\end{eqnarray}
\begin{large}
Martin Dohlus, Evgeni Saldin and Evgeni Schneidmiller
\end{large}

\textsl{\\Deutsches Elektronen-Synchrotron DESY, \\Notkestrasse
85, 22607 Hamburg, Germany}
\begin{eqnarray}
\nonumber
\end{eqnarray}
\begin{large}
Mikhail Yurkov
\end{large}

\textsl{\\Particle Physics Laboratory (LSVE), Joint Institute for
Nuclear Research, \\141980 Dubna, Moscow Region, Russia}

\newpage

\end{center}
\end{widetext}

 \maketitle

\section{\label{sec:intro}INTRODUCTION}

Nowadays the production of high-brightness electron bunches
constitutes one of the most challenging and interesting activities
for particle accelerator physicists.

The growing interest of the community in this kind of beams is
justified by important applications such as self-amplified
spontaneous emission (SASE)-free-electron-lasers operating in the
x-ray region (see, amongst others, \cite{XRAY}). Similar bunches
are also being considered for production of femtosecond radiation
pulses by simpler schemes based on Cherenkov and Transition
Radiation \cite{WALTER}.

One of the challenges faced by physicists involved in this area is
constituted by the presence of self-field induced collective
effects. These effects may spoil the brightness of the electron
beam.

Self-fields obey the usual Li\'{e}nard-Wiechert expressions; the
fields generated whenever an electron bunch undergoes a motion
under the influence of external forces are different from the
usual space-charge self-interaction, and are usually neglected in
cases in which one does not deal with high-brightness beams.

The equations for the field form, together with the dynamical
equations for the particle motion, a formidable self-consistent
problem. Its complete solution, that is the particle evolution, is
obtained only when one is able to solve simultaneously the
equations for the fields (electrodynamical problem) and the
equations of motion (dynamical problem): this task can be
performed with the help of self-consistent computer simulations.
An example of such codes is given (see \cite{ROTH}) by the program
$TraFiC^{4}$, which will be used throughout this paper.

$TraFiC^{4}$ has been used to predict self-field related effects
in the bunch compressor chicane to be used before the main linac
for the XFEL at DESY and it is, at the time being, the only fully
developed code used for XFEL modelling both at DESY and SLAC.
Results show that the projected transverse emittance of the bunch
grows from $0.8~mm~mrad$ after the injection to the significantly
enhanced value of $2.6~ mm~mrad$ after the compressor \cite{TDR}.

Transverse dynamics is addressed by the code in two steps: first,
the transverse electromagnetic forces, which are well defined and
measurable physical quantities, are calculated separately and,
second, the equation of motion is solved in a self-consistent way.

\begin{figure}
\includegraphics*[width=90mm]{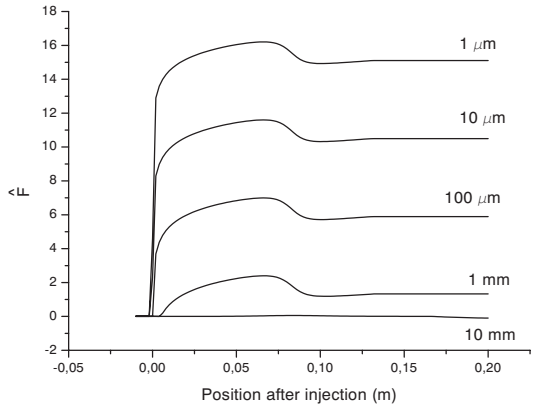}% Here is how to import EPS art
\caption{\label{FIG1} Radial force normalized to $e^2 \lambda/(4
\pi \varepsilon_0 R)$, $\hat F$, felt by a particle with different
vertical displacements $h$ positioned at the center of a line
bunch $200 ~\mu m$ long as the bunch enters a circle of radius
$R=1~m$. Here $\gamma = 100$. These simulations were obtained by
means of the code $TraFiC^4$.}
\end{figure}

From this viewpoint a thorough understanding of transverse
electromagnetic forces by themselves is of paramount importance
since it allows to gain confidence in full simulation results. The
behavior of these forces can be obtained numerically by using the
electromagnetic solver of the code for a given source distribution
evolving rigidly along a certain trajectory (i.e. a zero energy
spread is assumed throughout the bunch evolution), but numerical
results alone are not sufficient in order to reach a full
understanding of how electromagnetic forces act. Consider for
example the plot in Fig. \ref{FIG1}, obtained by using the
electromagnetic solver of $TraFiC^4$.

The figure shows the normalized radial force (i.e. the force
normalized to $e^2 \lambda/(4 \pi \varepsilon_0 R)$ in the
direction orthogonal to the test particle velocity and still lying
on the bending plane) felt by a test particle. Here $\lambda$ is
the bunch density while $h$ is the vertical (orthogonal to the
bending plane) displacement with respect to the center of a
one-dimensional bunch with rectangular density distribution
function.

The plot shows interesting characteristics: a very sharp feature
at the injection position, a transient and a steady state, which
we can observe but for which there is no intuitive physical
explanation. The electromagnetic problem is, of course, only a
part of the formidable evolution problem but its understanding is
$per se$ an important physical issue.

Analytical investigations of this simplified, although very
important situation are therefore needed in order to see the
meaning of the simulation results, besides providing, from a more
practical viewpoint, important cross-checks. We will show that
through this kind of study we can understand the features of Fig.
\ref{FIG1} from a physical viewpoint, deal with characteristic
times and lengths and, as a very practical result, validate the
simulation results by producing analytical cross-checks.

The self-interaction in the longitudinal direction (parallel, at
any time, to the velocity vector by definition) is responsible for
the energy exchange between the system and the acceleration field
and for all CSR (Coherent Synchrotron Radiation)-related
phenomena, which have been studied extensively elsewhere (see,
amongst others, \cite{ROTH}... \cite{BORL}).

Self-forces in the transverse direction were first addressed, in
the case of a circular motion, and from an electrodynamical
viewpoint, in \cite{TALM}. Further analysis (\cite{RUI2},
\cite{LEE}... \cite{STUP}) considers, again, the case of circular
motion both from an electrodynamical and a dynamical viewpoint. We
proposed a fully electrodynamical analysis of a bunch moving in an
arc of a circle in \cite{OURS}, where, for the reasons explained
before, we were not interested in the full evolution problem.
Nevertheless, in that paper we addressed the issue of
understanding the electromagnetic interactions only in the
framework of a 1D model. Before our study in \cite{OURS} there was
no explanation for the simulation results shown in Fig.
\ref{FIG1}. The work in \cite{OURS} provided a qualitative
explanation of these results: such qualitative character of the
explanation is due to the fact that the model in \cite{OURS} is
one-dimensional, and does not allow any quantitative investigation
when (as in Fig. \ref{FIG1}) a vertical displacement is
introduced. Still it was possible to demonstrate that the sharp
feature near the injection point is due to interactions of the
test particle with particles in front of it, while the transient
part is due to the interaction with particles behind it.

In this paper we aim at an extension of the model proposed in
\cite{OURS} which allows a fully electrodynamical analysis of a
bunch endowed with a vertical dimension. Since a bunch with
vertical extent can be always thought of as a superposition of
displaced charge lines, all the relevant physical aspects of the
problem are included in the study of a simpler model, constituted
by a one-dimensional line bunch and a test particle with vertical
displacement, which is the situation studied in Fig. \ref{FIG1}.
In this article, our explanation of the features in Fig.
\ref{FIG1} will be eventually refined to a quantitative level.

The paper is organized as follows. We first treat, in Section
\ref{sec:two}, the transverse interaction between two particles
moving in an arc of a circle, supposing that one of the two
particles has a vertical displacement $h$ with respect to the
source. By integration of our results we consider, in Section
\ref{sec:transientbunch}, a stepped-profile electron bunch
interacting with a test particle with vertical displacement
entering an arc of a circle and we discuss all the characteristic
lengths involved. Finally, in Section \ref{sec:concl}, we come to
a summary of the results obtained and to conclusions.

\section{\label{sec:two}TRANSVERSE INTERACTION BETWEEN TWO ELECTRONS}

We will first address the case of  two electrons moving in an arc
of a circle of radius $R$, in such a way that one of the two is
displaced, with respect to the reference trajectory, by a quantity
$h$ in the vertical (perpendicular to the bending plane)
direction. We will see that, as concerns the interaction in the
radial direction (in the bending plane) it does not matter which
particle is endowed with this displacement.

The electro-magnetic force which one of the two particles
(designated with "T", i.e. the test particle) feels, due to the
interaction with the other one (designated with "S", i.e. the
source particle), is given by

\begin{equation}
{\bm F}({\bm {r_{\mathrm{T}}}},t) = e{\bm E}({\bm
{r_{\mathrm{T}}}},t) + ec{\bm{\beta_{\mathrm{T}}}} \times {\bm
B}({\bm {r_{\mathrm{T}}}},t) \label{lorentz},
\end{equation}
where ${\bm {r_{\mathrm{T}}}}$ is the position of the test
particle, $e$ is the electron charge with its own (negative) sign,
$\bm{\beta_{\mathrm{T}}}$ is the velocity of the test particle
normalized to the speed of light, $c$, while $\bm{E}({\bm
{r_{\mathrm{T}}}},t)$ and $\bm{B}({\bm {r_{\mathrm{T}}}},t)$ are,
respectively, the electric and the magnetic field generated at a
given time $t$ by the source particle S, at the position of the
test particle T, namely

\begin{eqnarray} {\bm E}({\bm {r_{\mathrm{T}}}},t)={e \over {4 \pi
\varepsilon_\mathrm{0}}}\Biggl\{{1 \over \gamma_\mathrm{S}^{2}}
{{\bf {\hat{n}}} - {\bm {\beta_\mathrm{S}}}
\over{R_\mathrm{ST}^{2} \left({1-{\bf {\hat n}} \cdot {\bm
{\beta_\mathrm{S}}}}\right)^{3}}} + \nonumber\\&\cr + {1\over c}
{{\bf{\hat n}} \times \left[{\left({{\bf{\hat n}}-{\bm
{\beta_\mathrm{S}}}}\right)\times {\dot{\bm
\beta}_{\bm{\mathrm{S}}}}}\right] \over{R_\mathrm{ST} \left({1-
{\bf{\hat n}} \cdot {\bm {\beta_\mathrm{S}}}}\right)^{3}}}
\Biggr\}~ && \label{Efield}
\end{eqnarray}
and

\begin{equation}
{\bm B}({\bm {r_{\mathrm{T}}}}, t)={1\over c} {\bf{\hat n}}
\times {\bm E}({\bm {r_{\mathrm{T}}}},t)~. \label{Mfield}
\end{equation}
Here ${\bm{\beta_\mathrm{S}}}$ and ${\dot{\bm
\beta}_{\bm{\mathrm{S}}}}$ are, respectively, the dimensionless
velocity and its time derivative at the retarded time $t'$,
$R_\mathrm{ST}$ is the distance between the retarded position of
the source particle and the present position of the test electron,
$\bf{\hat{n}}$ is a unit vector along the line connecting those
two points and $\gamma_\mathrm{S}$ is the usual Lorentz factor
referred to the source particle at the retarded time $t'$.

In contrast to the case studied in \cite{OURS}, the transverse (by
definition, orthogonal to ${\bm{\beta_\mathrm{T}}}$) projection of
${\bm F}({\bm {r_{\mathrm{T}}}},t)$ has now components both in the
bending plane and perpendicular to it. We will introduce,
therefore, two unit vectors $\bf{\hat e_{h}}$, in the direction
perpendicular to the bending plane, and $\bf{\hat e_{p}}$, in the
radial direction, i.e. orthogonal to $\bm{\beta_\mathrm{T}}$ and
lying in the bending plane. Of course, the transverse component of
${\bm F}({\bm {r_{\mathrm{T}}}},t)$ can still be written,
following \cite{OURS}, as the sum of contributions from the
velocity ("C", Coulomb) and the acceleration ("R", Radiation)
fields, namely

\begin{equation}
{\bm {F_\bot}}({\bm {{r_{\mathrm{T}}}}},t) = {\bm {F_\mathrm{\bot
C}}}({\bm {r_{\mathrm{T}}}},t)+{\bm {F_\mathrm{\bot R}}}({\bm
{r_{\mathrm{T}}}},t), \label{transvlorentz}
\end{equation}
where

\begin{equation}
{\bm F_{\bm{\mathrm{\bot C}}}}({\bm {r_{\mathrm{T}}}},t) = {e^2
\over {4 \pi \varepsilon_\mathrm{0}}} {{\bf{n_\bot}}\left(1 -
{\bm{\beta_\mathrm{S}}}\cdot{\bm{\beta_\mathrm{T}}}\right) -
{\bm{\beta_\mathrm{\bot S}}}\left(1 - {\bf{\hat{n}}}\cdot
{\bm{\beta_\mathrm{T}}}\right)\over{\gamma_\mathrm{S}^2
R_\mathrm{ST}^2
\left(1-{\bf{\hat{n}}}\cdot{\bm{\beta_\mathrm{S}}}\right)^3}}
\label{Coulomb}
\end{equation}
and
\begin{widetext}
\begin{eqnarray}
{\bm F_{\bm{\mathrm{\bot R}}}}({\bm {r_{\mathrm{T}}}},t) = {e^2
\over {4 \pi \varepsilon_\mathrm{0}}} \times & \cr \times \Bigg[ {
{\bf{n_\bot}}
\left({\bf{\hat{n}}}\cdot{{\dot{\bm{\beta}}_{\bm{\mathrm{S}}}}}\right)\left(1
- {\bm{\beta_\mathrm{S}}}\cdot{\bm{\beta_\mathrm{T}}}\right)-
{\bm{\beta_\mathrm{\perp
S}}}\left({\bf{\hat{n}}}\cdot{{\dot{\bm{\beta}}_{\bm{\mathrm{S}}}}}\right)\left(1-
{\bf{\hat{n}}}\cdot{\bm{\beta_\mathrm{T}}}\right)\over{R_\mathrm{ST}
\left(1-{\bf{\hat{n}}}\cdot{\bm{\beta_\mathrm{S}}}\right)^3}} - {
\dot{\bm{\beta}}_{\bm{\mathrm{\perp T}}}\left(1 -
{\bf{\hat{n}}}\cdot{\bm{\beta_\mathrm{T}}}\right) +
{\bf{\hat{n}_\perp}}\left({\bm{\beta_\mathrm{T}}}\cdot{{\dot{\bm{\beta}}_{\bm
{\mathrm{S}}}}}\right) \over{R_\mathrm{ST}
\left(1-{\bf{\hat{n}}}\cdot{\bm{\beta_\mathrm{S}}}\right)^2}}\Bigg]\;
&. \label{Radiation}
\end{eqnarray}
\end{widetext}

\subsection{Tail-Head interaction: case of two particles in circular motion}

\begin{figure}[htb]
\includegraphics*[width=90mm]{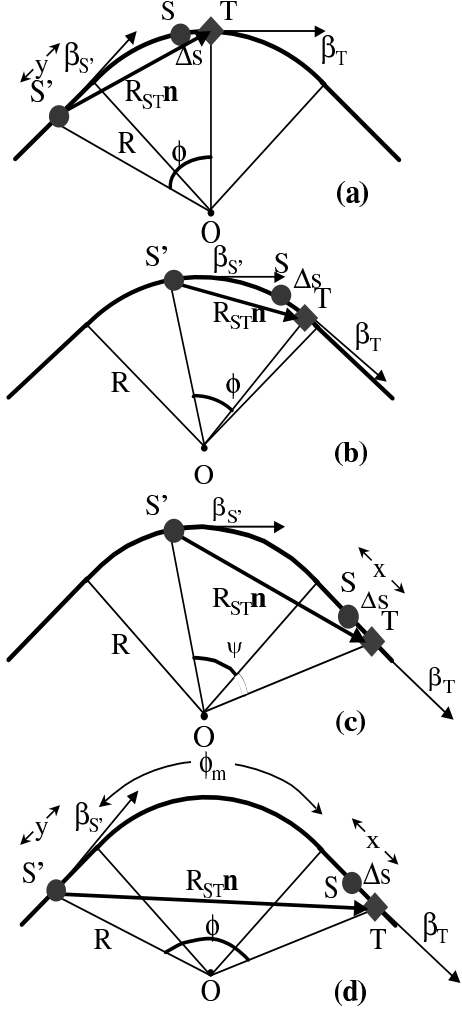}% Here is how to import EPS art
\caption{\label{FIG2} Relative configuration of the retarded
source point S' and the test point T for a system of two electron
passing a bending magnet. The test particle is understood to be
displaced by a quantity $h$ in the vertical direction. The figure
shows only the projection of the motion on the bending plane. }
\end{figure}

We will first consider the situation in which the test particle is
in front of the source. In this case, one can refer to Fig.
\ref{FIG2} for all the possible configurations of the present
position of the test electron and the retarded position of the
source with respect to the arc. The vertical displacement $h$ of
one of the two particles is to be imagined in the direction
perpendicular to the figure plane. Let us start with the steady
state case in Fig. \ref{FIG2}b. We can define with $\Delta s$ the
curvilinear distance between the present position of the test and
of the source particle; $\phi$ will indicate the angular distance
between the retarded position of the source and the present
position of the test electron, and it will be designated as the
retarded angle. Finally, $h$ will be the vertical displacement.

In the following we will assume $\beta_\mathrm{S} =
\beta_\mathrm{T} = \beta$.
%This assumption naturally leads to the
%assumption of a zero energy-spread when one considers the
%evolution of an electron bunch, and, as already discussed in Sec.
%\ref{sec:intro}, it is consistent with our choice of analyzing
%only the electrodynamical aspect of the problem.
Therefore we can write Eq. (\ref{Coulomb}) and Eq.
(\ref{Radiation}) in the following way:

\begin{widetext}
\begin{equation}
{\bm{F_\mathrm{\bot C}}} = {e^2 \over {4 \pi
\varepsilon_\mathrm{0}} \gamma^2} {{{\bf{\hat{e}_h}} \left[ h
(1-\beta^2\cos\phi)\right]+{\bf{\hat{e}_p}}\left[2 R
\sin^2(\phi/2)(1+\beta^2)-\beta\sin\phi(h^2+4R^2\sin^2(\phi/2))^{1/2}\right]}
\over{\left[(h^2+4R^2\sin^2(\phi/2))^{1/2}-2\beta
 R \sin(\phi/2)\cos(\phi/2)\right]^3}}~, \label{Coultrig}
\end{equation}
\end{widetext}
\begin{widetext}
\begin{eqnarray}
{\bm{F_\mathrm{\bot R}}} = {e^2 \beta^2 c \over {4 \pi
\varepsilon_\mathrm{0}}} \times && \cr \times \Bigg\{2
\sin^2(\phi/2) {{\bf{\hat{e}_h}}\left[h (1 -\beta^2
\cos\phi)\right]+{2 {\bf{\hat{e}_p}} \sin(\phi/2) \left[R
\sin(\phi/2)(1+\beta^2)-\beta \cos(\phi/2)(h^2+4 R^2
\sin^2(\phi/2))^{1/2}\right]}\over{\left[
(h^2+4R^2\sin^2(\phi/2))^{1/2}-2 \beta R
\sin(\phi/2)\cos(\phi/2)\right]^3}} - && \cr -
{{\bf{\hat{e}_h}}\left[h \beta
\sin\phi\right]+{\bf{\hat{e}_p}}\left[\beta R
\sin\phi-\cos\phi(h^2+4 R^2
\sin^2(\phi/2))^{1/2}\right]\over{R\left[
(h^2+4R^2\sin^2(\phi/2))^{1/2}-2 \beta R
\sin(\phi/2)\cos(\phi/2)\right]^2}}\Bigg\}~. && \label{Radtrig}
\end{eqnarray}
\end{widetext}
The retardation condition linking $\Delta s$, $h$ and $\phi$ reads

\begin{equation}
\Delta s = R\phi -\beta \left[h^2+4 R^2
\sin^2{\phi\over{2}}\right]^{1/2}~, \label{retsteady}
\end{equation}
As one can readily see by inspecting Eq. (\ref{retsteady}), when
we impose reasonable values for $\Delta s \ll R$ and $h \ll R$ we
obtain corresponding values of $\phi \ll 1$. We will therefore
assume $\phi \ll 1$ throughout this paper, and verify
$a~posteriori$ the validity of this assumption when studying
particular situations. Note that, by fixing $\phi \ll 1$ we keep
open the possibility of comparing $\phi$ with the synchrotron
radiation formation angle $1/\gamma$ (note that a deflection angle
smaller or larger than $1/\gamma$ is characteristic of the cases,
respectively, of undulator or synchrotron radiation). We can
therefore expand Eq. (\ref{Coultrig}) and Eq. (\ref{Radtrig}) to
the second non-vanishing order in $\phi$ thus obtaining

\begin{equation}
{\bm{F_\mathrm{\bot C}}} \simeq {e^2 \gamma^3 \over {4 \pi
\varepsilon_\mathrm{0} R^2}} {\bm{\Phi_\mathrm{C}}}(\hat{\phi})
\label{Coulexp}
\end{equation}
and

\begin{equation}
{\bm{F_\mathrm{\bot R}}} \simeq {e^2 \gamma^3 \over {4 \pi
\varepsilon_\mathrm{0} R^2}} {\bm{\Phi_\mathrm{R}}}(\hat{\phi})~,
\label{Radexp}
\end{equation}
where we define ${\bm{\Phi_\mathrm{C}}}$ and
${\bm{\Phi_\mathrm{R}}}$ as

\begin{equation}
{\bm{\Phi_\mathrm{C}}}(\hat{\phi}) = {{4{\bf{\hat{e}_h}}\left[\hat
h + 2 \hat
h/\hat{\phi}^2\right]+{\bf{\hat{e}_p}}\left[\hat{\phi}^2-4\hat{h}^2/\hat{\phi}^2
\right]}\over {\hat{\phi} \left[1+\hat{\phi}^2/4+
\hat{h}^2/\hat{\phi}^2\right]^3}} \label{PhiC}
\end{equation}
and

\begin{equation}
{\bm{\Phi_\mathrm{R}}}(\hat{\phi}) =
{{{\bf{\hat{e}_h}}\hat{\phi}^2\hat h\left[1 -4 \hat
{h}^2/\hat{\phi}^4\right]+{\bf{\hat{e}_p}}\left[2-\hat{\phi}^2-3
\hat{h}^2+4\hat{h}^2/\hat{\phi}^2 \right]}\over {\hat{\phi}
\left[1+\hat{\phi}^2/4+ \hat{h}^2/\hat{\phi}^2\right]^3}}
\label{PhiR}
\end{equation}
Here and above $\hat{\phi} = \gamma \phi$. This normalization
choice, already treated in \cite{SALLONG}, is quite natural,
$1/\gamma$ being the synchrotron radiation formation-angle at the
critical wavelength. In the derivation of Eq. (\ref{Coulexp}) and
Eq. (\ref{Radexp}) (and in the following, too) we understood $\hat
\phi \gg 1/\gamma$, which is justified by the ultrarelativistic
approximation. Moreover we defined $\hat h = h \gamma^2/R$;  in
order to understand this definition we first write down the
retardation condition Eq. (\ref{retsteady}) in approximate form.
Since we are already working in the limit $\Delta s/R \ll 1$ and
$\phi \ll 1$, then in order to have $\Delta s > 0$, $h < R \phi$
must be satisfied. We will automatically recover from our results
that the assumption $h \ll R \phi$ is sufficiently good for our
purposes. Therefore Eq. (\ref{retsteady}) can be approximated
with:

\begin{equation}
\Delta s = (1-\beta)R\phi + {R\phi^3\over{24}} - {\beta h^2\over{2
R \phi}}~. \label{retsteadyappr}
\end{equation}
The latter can be written down in dimensionless form too as

\begin{equation}
\Delta \hat{s} = {\hat\phi\over{2}} + {\hat{\phi}^3\over{24}} - {
\hat{h}^2\over{2 \hat\phi}}~, \label{retsteadydimlessappr}
\end{equation}
which explains the choice of the definition for $\hat h$.

The reader may easily check that Eq. (\ref{PhiC}), Eq.
(\ref{PhiR}) and Eq. (\ref{retsteadydimlessappr}) (as well as Eq.
(\ref{Coultrig}) and Eq. (\ref{Radtrig})) are generalizations of
the expressions given in \cite{OURS} by taking their limit for
$\hat h \longrightarrow 0$.

The following expression, which is valid for the total transverse
force felt by the test particle can be then trivially derived
\begin{equation}
{\bm{F_\mathrm{\bot}}} \simeq {e^2 \gamma^3 \over {4 \pi
\varepsilon_\mathrm{0} R^2}} {\bm\Phi}(\hat{\phi})~,
\label{Totexp}
\end{equation}
where $\Phi$ is defined by

\begin{equation}
{\bm\Phi}(\hat{\phi}) =
{\bm{\Phi_\mathrm{R}}}(\hat{\phi})+{\bm{\Phi_\mathrm{C}}}(\hat{\phi}),\label{Phi}
\end{equation}

\subsection{Tail-Head interaction: case (a)}

Let us now consider the other cases depicted in Fig. \ref{FIG2}a,
c and d. While the case in Fig. \ref{FIG2}b deals with the steady
state situation in which the present position of the test and the
retarded position of the source are both in the bend, Fig.
\ref{FIG2}a, c and d deal with transient situations in which we
can find the retarded source and the present test in the straight
line before and after the bend too.

Consider the situation in Fig. \ref{FIG2}a. In this case, under
the assumption $h \ll (y+R\phi)$, the retardation condition reads

\begin{equation}
\Delta \hat{s} \simeq {\hat y + \hat \phi \over{2}} - {1\over{2}}
{\hat{h}^2-\hat{\phi}^3/3(\hat y + \hat \phi/4)\over{\hat y + \hat
\phi}}~, \label{retcondA}
\end{equation}
where we introduced the normalized quantity $\hat{y} = y
\gamma/R$, which is  just $y/R$ normalized to the synchrotron
radiation formation angle, $1/\gamma$.

In this case the source particle is only responsible for a
velocity field contribution, therefore ${\bm F_\mathrm{\bot}} =
{\bm F_\mathrm{\bot C}}$. By direct use of Eq. (\ref{Coulomb}),
one can find the exact expression for ${\bm F_\mathrm{\bot}}$

\begin{widetext}
\begin{equation}
{\bm F_\mathrm{\bot}} = {e^2  \over {4 \pi \varepsilon_\mathrm{0}
\gamma^2}} {{{\bf{\hat{e}_h}} \left\{h [1-\beta^2 \cos\phi]
\right\}+ {\bf{\hat{e}_p}}\left\{ -R_{ST} \beta \sin\phi +\left[ R
(1+\beta^2)(1-\cos\phi)+y\sin\phi\right]\right\}}
\over{\left\{R_{ST}-\beta(y+R \sin\phi)\right\}^3}} \label{triga}
\end{equation}
\end{widetext}
$R_{ST}$ being given by

\begin{equation}
R_{ST} = \left[(y+R \sin
\phi)^2+R^2(1-\cos\phi)^2+h^2\right]^{1/2}~. \label{RST}
\end{equation}

\begin{figure}
\includegraphics*[width=90mm]{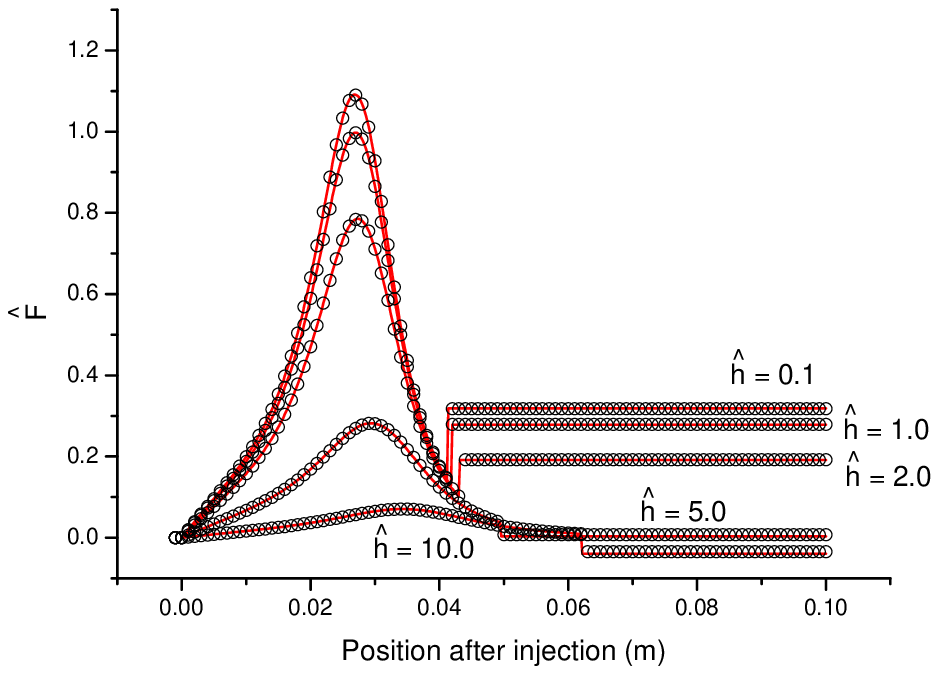}% Here is how to import EPS art
\caption{\label{FIG3} Normalized radial force $\hat{F}$ for a
two-particle system entering a hard-edge bending magnet as a
function of the position after injection. Results for different
values of $\hat h$ are shown. The solid lines show analytical
results; the circles describe the outcome from
$\mathrm{TRAFIC}^4$. Here $\Delta \hat{s} = 5.0$. }
\end{figure}

Expanding the trigonometric functions in Eq. (\ref{triga}) and
using normalized quantities one finds:

\begin{widetext}
\begin{equation}
{\bm F_\mathrm{\bot}} \simeq { e^2  \over {4 \pi
\varepsilon_\mathrm{0} }} {4 \gamma^3 \over
{R^2}}(\hat{y}+\hat{\phi})^2 {\left\{ {\bf{\hat{e}_h}}\left[\hat h
(\hat y + \hat \phi)(2+\hat{\phi}^2)\right]+{\bf{\hat{e}_p}}\left[
\hat{y}^2\phi
+\hat{y}(\hat{\phi}^2+\hat{\phi}^4/2)+{\hat{\phi}^5/4}-\hat{h}^2\phi
\right]
\right\}\over{\left[(\hat{y}+\hat{\phi})^2+{\hat{\phi}^4/4}+\hat{h}^2\right]^3}}
\label{expa}
\end{equation}
\end{widetext}
It can be easily verified that, as it must be, Eq. (\ref{expa})
reduces to Eq. (\ref{Coulexp}) in the limit $\hat y \rightarrow
0$. Also, the reader may check that in the limit $\hat h
\rightarrow 0$, Eq. (\ref{expa}) reduces to already derived
expressions in \cite{OURS}.

Let us now define the normalized radial force $\hat{F} = F_p /
[e^2/(4\pi\varepsilon_0 R \Delta s)]$. It is possible, by means of
Eq. (\ref{expa}), to plot $\hat{F}$ as a function of the position
after the injection (defined by the entrance of the test particle
in the hard-edge magnet) for a fixed value of $\Delta \hat{s} =
\Delta s \gamma^3/R = 5.0$ and different values of $h$. In Fig.
\ref{FIG3} we compare such a plot with numerical results from the
code $\mathrm{TRAFIC}^4$ (see \cite{ROTH}).

Note that, as already pointed out in \cite{OURS}, at the position
which corresponds to the entrance of the retarded source in the
magnet there is a discontinuity in the plots. This is linked to
our model choice, and it is due to the abrupt (hard edge magnet)
switching on of the acceleration fields.

As general remark to Fig. \ref{FIG3} and, in fact to Fig.
\ref{FIG4}, \ref{FIG5} and Fig.\ref{FIG11}, the perfect agreement
 between our calculations and numerical results by
$\mathrm{TRAFIC}^4$ provides, \textit{per se}, an excellent
cross-check between our analytical results (with their assumptions
and applicability region) and simulations.

\subsection{Tail-Head interaction: case (c)}

We will now move to the case depicted in Fig. \ref{FIG2}c, in
which the source particle has its retarded position inside the
bend and the test particle has its present position in the
straight line following the magnet. We will define with $x$ the
distance, along the straight line after the magnet, between the
end of the bend and the present position of the test particle.
Here $\hat{\phi}_\mathrm{m} = \gamma \phi_\mathrm{m}$,
$\phi_\mathrm{m}$ being the angular extension of magnet, and
$\hat{x}=\gamma x/R$, the reason for this normalization choice for
$x$ being identical to that for $y$.

The retardation condition reads

\begin{equation}
\Delta \hat{s} \simeq
{\hat{\psi}+\hat{x}\over{2}}+{\hat{\psi}^3\over{24}}{\hat{\psi}+
4\hat{x}\over{\hat{\psi}+\hat{x}}}-{\hat {h}^2\over{2
(\hat{\psi}+\hat{x})}}. \label{retcond:b}
\end{equation}
In contrast with the case of Fig. \ref{FIG2}a, here we have
contributions from both velocity and acceleration field. Again, by
direct use of Eq. (\ref{Coulomb}) and Eq. (\ref{Radiation}) one
can find the exact expression for the transverse electromagnetic
force exerted by the source particle on the test particle

\begin{equation}
{\bm F_\mathrm{\bot}} = {\bm F_\mathrm{\bot C}} + {\bm
F_\mathrm{\bot R}}~, \label{Fperptotb}
\end{equation}
where
\begin{widetext}
\begin{equation}
{\bm F_\mathrm{\bot C}} = { e^2  \over {4 \pi
\varepsilon_\mathrm{0} \gamma^2}} {{\bf{\hat{e}_h}} \left[
(1-\beta^2 \cos \psi)h\right] + {\bf{\hat{e}_p}} \left[ R
(1-\cos\psi)(1-\beta^2 \cos\psi)-\beta\sin\psi \left(R_{ST}-\beta
x -\beta R\sin\psi \right) \right] \over{\left[R_{ST}-\beta x
\cos\psi -\beta R \sin\psi \right]^3}}~ \label{FperpbC}
\end{equation}
and
\begin{eqnarray}
{\bm F_\mathrm{\bot R}} = { e^2 \over {4 \pi
\varepsilon_\mathrm{0} R }} \Bigg\{\Bigg[{{{\bf{\hat{e}_h}}
\left[\beta^2
h(1-\beta^2\cos\psi)(R+x\sin\psi-R\cos\psi)\right]}\over{\left[R_{ST}-\beta
x \cos\psi -\beta R \sin\psi \right]^3}}+&& \cr +
{{{\bf{\hat{e}_p}} \beta^2 \left[(R+x\sin\psi-R\cos
\psi)(R(1-\cos\psi)(1-\beta^2 \cos\psi)- \beta\sin
\psi(R_{ST}-\beta x + \beta R \sin\psi ))\right]}
\over{\left[R_{ST}-\beta x \cos\psi -\beta R \sin\psi
\right]^3}}\Bigg]-&&\cr- {{\bf{\hat{e}_h}}\beta^2 \left[h
\sin\psi\right]+ {\bf{\hat{e}_p}}
\left[R(1-\cos\psi)\beta^2\sin\psi-\beta\cos\psi( R_{ST}-\beta
x-\beta R\sin\psi) \right] \over{\left[R_{ST}-\beta x \cos\psi
-\beta R \sin\psi \right]^2}}\Bigg\}&&~.\label{FperpbR}
\end{eqnarray}
\end{widetext}
$R_{ST}$ being now
\begin{equation}
R_{ST}=\left((x + R\sin\psi)^2+R^2(1-\cos\psi)^2 +h^2\right)^{1/2}
\label{RSTc}
\end{equation}

Expanding the trigonometric functions in Eq. (\ref{FperpbC}) and
Eq. (\ref{FperpbR}), and using normalized quantities one finds:

\begin{figure}
\includegraphics[width=90mm]{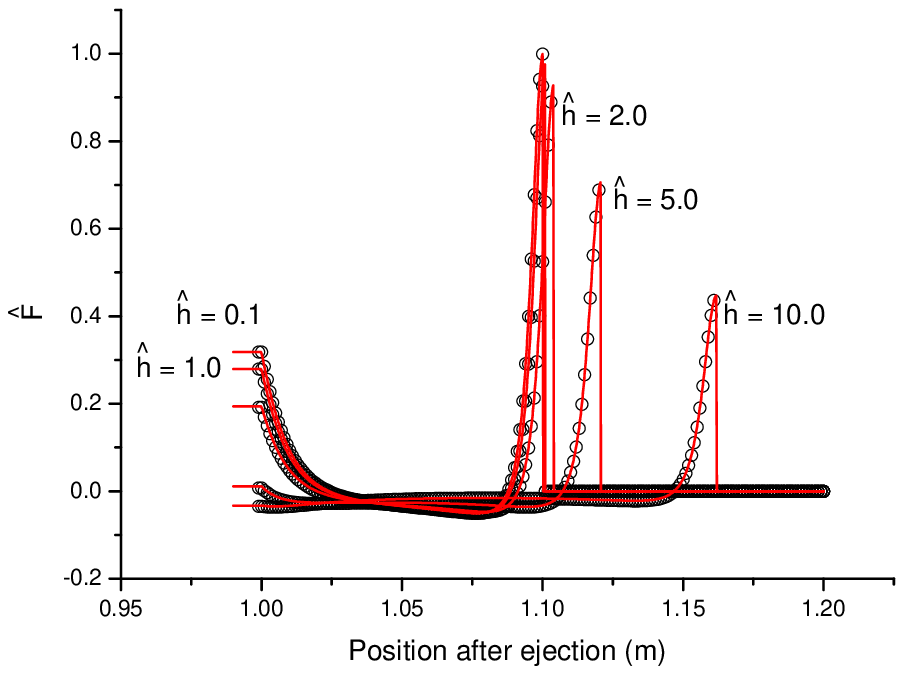}% Here is how to import EPS art
\caption{\label{FIG4} Normalized radial force $\hat{F}$ for a
two-particle system leaving a hard-edge bending magnet as a
function of the position after the ejection. Results for different
values of $\hat h$ are shown. The solid lines show analytical
results; the circles describe the outcome from
$\mathrm{TRAFIC}^4$. Here $\Delta \hat{s} = 5.0$.}
\end{figure}
\begin{widetext}
\begin{equation}
{\bm F_\mathrm{\bot C}} = {2 e^2 \gamma^3 \over {4 \pi
\varepsilon_\mathrm{0} R^2}} { 2{\bf{\hat{e}_h}} \left[\hat h
(2+\hat{\psi}^2)(\hat x +\hat \psi)^3 \right]+ {\bf{\hat{e}_p}}
\hat \psi (\hat x+\hat \psi)^2 \left[ -2\hat{x}^2
+\hat{x}\left(\hat{\psi}^3 -2\hat{\psi}\right)+{\hat{\psi}^4/2} -
2 \hat{h}^2 \right] \over{\left[(\hat{x}+ \hat{\psi})^2 +
({\hat{\psi}^2/4})(2\hat{x}+\hat{\psi})^2 +\hat{h}^2 \right]^3 }}
~, \label{apprFbC}
\end{equation}
and

\begin{eqnarray}
{\bm F_\mathrm{\bot R}} = {2 e^2 \gamma^3 \over {4 \pi
\varepsilon_\mathrm{0} R^2}} (\hat{x}+ \hat{\psi})\Bigg\{ { -2
{\bf{\hat{e}_h}}\left[ \hat h \hat \psi (\hat x+\hat \psi)
\right]+ {\bf{\hat{e}_p}} \left[
\hat{x}^2+\hat{x}\hat{\psi}(2-\hat{\psi}^2)+\hat{\psi}^2- (3/4)
\hat{\psi}^4 +\hat{h}^2\right] \over{\left[(\hat{x} +
\hat{\psi})^2 + (\hat{\psi}^2/4)(2\hat{x}+\hat{\psi})^2 +\hat{h}^2
\right]^2}} +&&\cr+ {2{{\bf{\hat{e}_h}}
(\hat{x}+\hat{\psi})^2(2+\hat{\psi}^2)(\hat{x}+\hat{\psi}/2)\hat{\psi}
\hat{h} +{\bf{\hat{e}_p}}(\hat{x}+\hat{\psi})
(\hat{x}+\hat{\psi}/2)\hat{\psi}^2
\big[-2\hat{h}^2-2\hat{x}^2+\hat{x}\hat{\psi}(-2+\hat{\psi^2})
+\hat{\psi}^4/2\big] }\over{\left[(\hat{x} + \hat{\psi})^2 +
(\hat{\psi}^2/4)(2\hat{x}+\hat{\psi})^2 +\hat{h}^2
\right]^3}}\Bigg\}&&\label{apprFbR}
\end{eqnarray}
\end{widetext}
Similarly to the latter case, it can be easily verified that Eq.
(\ref{apprFbC}) and Eq. (\ref{apprFbR}) reduce to Eq.
(\ref{Coulexp}) and Eq. (\ref{Radexp}), respectively,  in the
limit $x \rightarrow 0$. Moreover, in the limit $h \rightarrow 0$,
they reduce to already known expressions given in \cite{OURS}.
Again, one can plot the normalized radial force $\hat{F}$ as a
function of the position after the ejection, defined by the exit
of the test particle from the hard-edge magnet, for different
values of $\hat h$ and a fixed value of $\Delta \hat{s} = \Delta s
\gamma^3/R=5.0$. In Fig. \ref{FIG4} we compare such a plot with
numerical results from $\mathrm{TRAFIC}^4$.

At the position which corresponds to the exit of the retarded
source from the magnet there is a discontinuity in the plots.
This, again, is linked to our model choice, and it is due to the
fact that, for particles on axis, after the retarded source has
left the magnet there is only Coulomb repulsion along the
longitudinal direction.

\begin{figure}
\includegraphics[width=90mm]{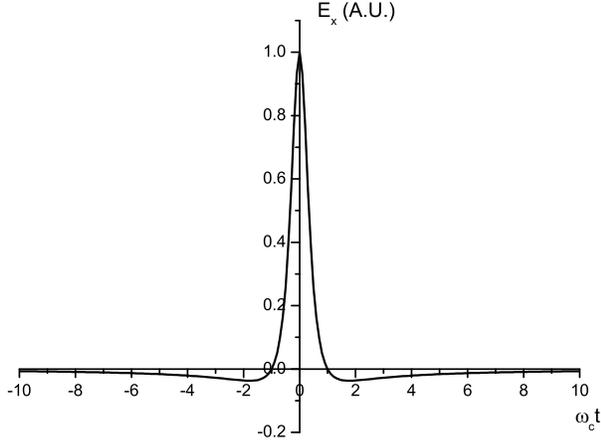}% Here is how to import EPS art
\caption{Time variation of a synchrotron radiation pulse generated
by a highly relativistic electron moving in a circle as seen by an
observer in the orbital plane \label{FIGSYN}}
\end{figure}

It is suggestive to notice the resemblance of the peaks shown in
Fig. \ref{FIG4} with half of the time pulse of the radial electric
field from usual synchrotron radiation process (see Fig.
\ref{FIGSYN}, and \cite{DIAGN}). This is not a coincidence. The
test particle is, indeed, far away from the source with respect to
the formation length $R/\gamma^3$ and the magnetic field
contribution to the Lorentz force can be expected to have the same
behavior of the electric field contribution, since ${\bm
B}=({\bf{\hat n}} \times {\bm E})/c$. The only difference is that
the observer is now "running away" from the electromagnetic signal
which will result in a spreading of the pulse of about a factor
$(1-\beta)^{-1}$. Since $R=1~ m$ and $\gamma = 100$ we expect the
pulse to be long about $(R/\gamma^3) (1-\beta)^{-1} \sim
10^{-2}~m$.

\subsection{Tail-Head interaction: case (d)}

The last Tail-Head case left to discuss is depicted in Fig.
\ref{FIG2}d; the source particle has its retarded position in the
straight line before the bend, and the test particle has its
present position in the straight line following the magnet. The
retardation condition reads

\begin{eqnarray}
\Delta \hat{s} \simeq
{\hat{\phi}_\mathrm{m}+\hat{x}+\hat{y}\over{2}}+ && \cr +
{\hat{\phi}_\mathrm{m}^3/{24} \left[\hat{\phi}_\mathrm{m}+
4(\hat{x}+\hat{y})+12\hat{x}\hat{y}/\hat{\phi}_\mathrm{m}
\right]-\hat{h}^2/2\over{\hat{\phi}_\mathrm{m}+\hat{x}+\hat{y}}}&&.
\label{retcond:c}
\end{eqnarray}
In this case we have only velocity contributions. The exact
expression for the electromagnetic transverse force on the test
particle is

\begin{widetext}
\begin{equation}
{\bm F_\mathrm{\bot}} = {e^2 \over {4 \pi \varepsilon_\mathrm{0}
\gamma^2}} {{\bf{\hat{e}_h}}\left[\sin\theta(1-\beta^2
\cos\phi_m)\right]+{\bf{\hat{e}_p}}\left[\sin\delta
\cos\theta(1-\beta^2
\cos\phi_m)-\beta\sin\phi_m(1-\cos\delta\cos\theta)\right]\over{R_{ST}^2\left[
1-\beta\sin\phi_m\sin\delta\cos\theta-\beta\cos\phi_m\cos\delta\cos
\theta\right]^3}} \label{Fperpc}
\end{equation}
\end{widetext}
where
\begin{equation}
\sin\theta={h\over{R_{ST}}}~, \label{senteta}
\end{equation}
\begin{equation}
\cos\theta={(R_{ST}^2-h^2)^{1/2}\over{R_{ST}}}~, \label{costeta}
\end{equation}
\begin{equation}
\sin\delta={R+y\sin\phi_m-R\cos\phi_m\over{R_{ST}\cos\theta}}~,
\label{sendelta}
\end{equation}
and
\begin{equation}
\cos\delta={x+y\cos\phi_m+R\sin\phi_m\over{R_{ST}\cos\theta}}~,
\label{cosdelta}
\end{equation}
where $R_{ST}$ can be retrieved by the latter two equations and
some trivial trigonometry.

\begin{figure}
\includegraphics[width=90mm]{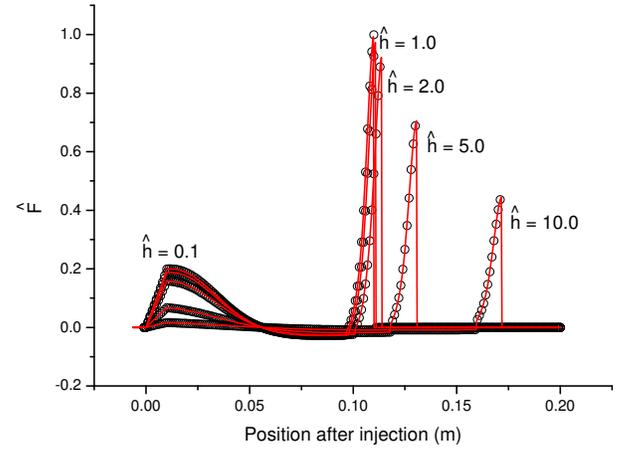}% Here is how to import EPS art
\caption{\label{FIG5} Normalized radial force $\hat{F}$ for a
two-particle system crossing a hard-edge bending magnet as a
function of the position of the test particle inside the magnet in
the case of a short magnet ${\phi}_\mathrm{m} \ll 1$. Results are
shown for different values of $\hat h$. The solid lines show
analytical results; the circles describe the outcome from
$\mathrm{TRAFIC}^4$. Here the normalized distance between the two
particles is $\Delta \hat{s} = 5$, while $\hat \phi_m = 1$.}
\end{figure}

Once again, expanding the trigonometric functions in Eqs.
(\ref{Fperpc}) ... (\ref{sendelta}) and using normalized
quantities one finds the following approximated expression for
${\bm F_\mathrm{\bot}}$:
\begin{widetext}
\begin{eqnarray}
{\bm F_\mathrm{\bot}} \simeq {e^2 \over {4 \pi
\varepsilon_\mathrm{0} R^2}} 4 \gamma^3
(\hat{x}+\hat{y}+\hat{\phi}_\mathrm{m})^2\hat
\phi_\mathrm{m}\times \nonumber\\&&\cr\times
{{\bf{\hat{e}_h}}\left[({2/{\hat\phi_m}}+{\hat\phi_m})(\hat x+
\hat y+\hat\phi_m)\hat h\right]+ {\bf{\hat{e}_p}}\left[
-{\hat{x}^2}+ {\hat{y}^2} +
({\hat{\phi}_\mathrm{m}^2})\hat{x}\hat{y}
+{\hat{x}(\hat{\phi}_\mathrm{m}^3/2 -\hat{\phi}_\mathrm{m})}
+{\hat{y}(\hat{\phi}_\mathrm{m}^3/2+\hat{\phi}_\mathrm{m})} +
{\hat{\phi}_\mathrm{m}^4/4}-\hat h^2\right]\over {\left\{
(\hat{x}+\hat{y}+\hat{\phi}_\mathrm{m})\left[\hat{x}
(1+\hat{\phi}^2_\mathrm{m}) +
\hat{y}+\hat{\phi}_\mathrm{m}+{\hat{\phi}^3_\mathrm{m}
/3}\right]-(\hat{\phi}^2_\mathrm{m} /12)[12
\hat{x}\hat{y}+4(\hat{x}+\hat{y})\hat{\phi}_\mathrm{m} +
\hat{\phi}_\mathrm{m}^2]+\hat
h^2\right\}^{3}}}~.&&~\label{Fperpcappr}
\end{eqnarray}
\end{widetext}
It is easy to verify that Eq. (\ref{Fperpcappr}) reduces,
respectively, to the steady state (Eq. (\ref{Coulexp})) when $x=0$
and $y=0$, to the transient case in Fig. \ref{FIG2}a when $x=0$
(Eq. (\ref{expa})) and to the transient case in Fig. \ref{FIG2}c
when $y=0$ (Eq. (\ref{FperpbC})). Moreover the reader may verify
that, in the limit $h\longrightarrow 0$, Eq. (\ref{Fperpcappr})
reduces to already known results in \cite{OURS}. Following the
treatment of the transient situations in Fig. \ref{FIG2}a and in
Fig. \ref{FIG2}c we plot, for this case too, a normalized
expression for the transient force, i.e. the usual $\hat{F}$, as a
function of the curvilinear position of the test particle ($s = 0$
indicates the entrance of the magnet) for different values of
$\hat h$ and a fixed value of $\Delta \hat{s} = \Delta s
\gamma^3/R$ and magnet length. In Fig. \ref{FIG5}, we compare our
analytical results with numerical results from
$\mathrm{TRAFIC}^4$, for a fixed value of $\Delta \hat{s} = 5.0$
and $\hat \phi_m= 1$ and for different values of the vertical
displacement $\hat h$.

\subsection{Head-Tail interaction}

Finally, we can deal with the situation in which the source
particle is ahead of the test electron, i.e. $\Delta s < 0$; we
will talk, in this case, of head-tail interaction. On the one hand
it is evident that, when $\Delta s < 0$,  the source particle is
ahead of the test electron at any time; on the other hand it is
not true that the present position of the test particle is, in
general, ahead of the retarded position of the test particle. As
already suggested in \cite{OURS}, if $\Delta s < 0$ and,
approximatively, $|\Delta s| < h$ the test particle overtakes the
retarded position of the source before the electromagnetic signal
reaches it. In this case, although we may still talk about
head-tail interaction, since $\Delta s < 0$, its real character is
very much similar to the case $\Delta s > 0$, just treated in the
previous subsections, in which the electromagnetic signal has to
catch up with the test particle. In order to understand the
physics involved in this situation it is sufficient to study the
cases $\Delta s < 0$ and $|\Delta s| > h$. The case $\Delta s < 0$
with $|\Delta s| < h$ will be treated from a qualitative viewpoint
only: its quantitative analysis, which may be interesting to
perform for the sake of completeness, presents stronger
mathematical difficulties and is left for future study. Anyway a
qualitative treatment of this situation is enough to reach a full
understanding of the interaction physics, which is our goal here.

Consider therefore the case $\Delta s < 0$ with $|\Delta s| > h$.
It is important to note that, once the steady state case is
studied, the situation in which the source particle is ahead of
the test electron can be treated immediately for all three (a, c
and d) transient cases in Fig. \ref{FIG2} (of course, with respect
to the figure, test and source particle exchange roles) on the
basis of the steady state case alone. In fact in that situation,
as it has already been said in \cite{OURS}, we can assume the
retarded angle $\phi$ small enough (the test particle "runs
against" the electromagnetic signal) that the actual trajectory
followed by the particles is not essential and one can use the
steady state expression to describe also the transient cases. It
can be shown that the only important contribution from the source
particle comes from the acceleration part of the Lienard-Wiechert
fields. Within our approximations, the only non-negligible
contribution is present in the situation (again, with the roles of
test and source particle inverted) depicted in Fig. \ref{FIG2}a
and Fig. \ref{FIG2}b, the latter just being the steady state case.

Let us deal with the steady state case of head-tail interaction in
the case $|\Delta s| > h$. As already discussed in \cite{OURS},
the difference with respect to the situation in which the test
electron is in front of the source is that the test electron "runs
against" the electromagnetic signal emitted by the source, while
in the other case it just "runs away" from it. Therefore the
relative velocity between the signal and the test electron is
equal to $(1+\beta)c$, instead of $(1-\beta)c$ as in the other
situation. Hence the retardation condition reads

\begin{equation}
\Delta s \simeq  R\phi + \beta \left[ R^2\phi^2 + h^2
\right]^{1/2} \label{retcondhtexact}
\end{equation}
or, solved for $\hat \phi$,

\begin{equation}
\hat\phi \simeq \Delta \hat{s}-\left[\Delta \hat{s}^2\left(1 -
{1\over{\gamma^2}}\right)-\hat h^2\right]^{1/2}~.
\label{retcondht}
\end{equation}
Note that, in the asymptotic for $\Delta \hat{s} \gg \gamma \hat
h$ we recover the result in \cite{OURS} for the one-dimensional
case.

In the general case, $\bm{\beta_\mathrm{S}}$ is almost parallel to
(and equal to) $\bm{\beta_\mathrm{T}}$ and antiparallel to the
projection of $\bf{\hat{n}}$ on the bending plane: it turns out
that the only important contribution to the radial force on the
orbital plane is given by the second term on the right side of Eq.
(\ref{Radiation}), and it is easy to check that

\begin{equation}
{F_\mathrm{\bot}} \simeq {e^2  \over {4 \pi
\varepsilon_\mathrm{0} R \Delta s}}~.
 \label{Fperpht}
\end{equation}
It may be worthwhile to underline that the force in Eq.
(\ref{Fperpht}) is, structurally, identical to the result in
\cite{OURS} although, of course, $\Delta s$ is now a function of
$h$ .

\section{\label{sec:transientbunch}TRANSVERSE INTERACTION BETWEEN AN
ELECTRON AND A BUNCH ENTERING A BEND FROM A STRAIGHT PATH}

In the previous Section we dealt with all the possible
configurations for a two-particle system moving in an arc of a
circle. Now we are ready to provide a quantitative explanation of
Fig. \ref{FIG1}.

After the discussion, in Section \ref{sec:two}, about head-tail
interactions within a system with the test particle behind the
source electron, one is led to the qualitative conclusion that,
within an electron bunch, interactions between sources in front of
the test particle and the test particle itself are important and,
in general, they must be responsible, at the entrance and at the
exit of the bending magnet, for sharp changes in the transverse
forces acting on the test electron. The quantitative change
depends, of course, on the position of the test particle inside
the bunch: the extreme cases are for the test particle at the head
of the bunch, where there are just interactions with electrons
behind the test particle (no head-tail interactions), and for the
test particle before the tail of the bunch, at a distance $|\Delta
s| > h$, where all the sources are in front of it (only head-tail
interactions in the regime $|\Delta s| > h$).

\subsection{Head-Tail interaction}

Consider now the situation in which a one-dimensional line bunch
interacts with a particle positioned at the center of the line but
vertically displaced by a quantity $h$. As already said we will
discuss, analytically, the head-tail interaction for $|\Delta
s|>h$ only. Since the trajectory followed by the bunch is not
important we can simply integrate Eq. (\ref{Fperpht}) and find the
contribution

\begin{widetext}
\begin{equation}
{F_\mathrm{p~HT}^\mathrm{B}}(\phi) \simeq  \Bigg\{
\begin{array}{c}
0~~~~~~~~~~~~~~~~~~~~~~~~~~~~~~~~~~~~~~~~~~~~~~~~~~~~~~~~~~~~\phi<0\\
e^2 \lambda_0 /(4 \pi \varepsilon_\mathrm{0} R) ~ \ln\left[\Delta
s_\mathrm{max}/(\Delta s_\mathrm{max} - R\phi(1+\beta))\right]
~~~~~~~~0<\phi< \Phi(h, \Delta s_\mathrm{max})\\
e^2 \lambda_0 /(4 \pi \varepsilon_\mathrm{0} R) ~ \ln[\Delta
s_\mathrm{max}/h]~~~~~~~~~~~~~~~~~~~~~~~~~~~~~~~~~~~~\phi>\Phi(h,\beta
h )
\end{array}~, \label{HTbunch}
\end{equation}
\end{widetext}
where "HT" stands for "head-tail" and $\Phi(h, \Delta s)$ is the
solution, in $\phi$, of Eq. (\ref{retcondhtexact}) at vertical
displacement $h$ and longitudinal distance $\Delta s$.

When a bunch longer than the vertical displacement $h$ enters the
magnet, the particles in front of  the bunch will interact with
the test electron following Eq. (\ref{HTbunch}), which models the
radial interaction on the basis of Eq. (\ref{Fperpht}).
Nevertheless, Eq. (\ref{Fperpht}) cannot describe the situation
when the sources at a distance shorter than the vertical
displacement $h$ begin to interact with the test electron. As a
result the head-tail interaction has two characteristic formation
lengths. The first one indicates the distance that the test
electron travels from the moment it is reached by the
electromagnetic signal from the first particle entering the bend,
till the moment it is reached by the electromagnetic signal
emitted as the particle at $\Delta s = -h$ enters the bend. This
is given by $L_1 = \beta c R (\phi_2-\phi_1) = \beta c R \phi_2$,
$\phi_2$ and $\phi_1$ being the solution of Eq. (\ref{retcondht})
when $\Delta s = l_b/2$ and $\Delta s = \beta h$, respectively
(the reader will recognize that $\phi_1 = 0$). In the limit for
$h<<l_b/2$ Eq. (\ref{retcondht}), substituted in the expression
for $L_1$, gives the rule of thumb $L_1 \simeq l_b/4$. The second
characteristic length is given by the distance that the test
electron travels from the moment it is reached by the
electromagnetic signal from the particle at $\Delta s = -h$ till
the moment it is reached by the electromagnetic signal emitted by
the particle with $\Delta s = 0$ as it enters the bend. This can
be estimated to be, roughly, equal to $\gamma h$, at least when
$h$ is not too large. In fact, in order to know the present
angular position of the test particle when $\Delta s = 0$ one
should solve the retardation condition in $\phi$

\begin{equation}
R \phi = \beta \left[h^2+4 R^2 \sin^2(\phi/2)\right]^{1/2}
\label{RETCONDH}
\end{equation}
In the limit in which $\sin(\phi/2)\simeq \phi/2$  we have $\phi
\simeq \gamma h/R$ and $L_2 \simeq \gamma h$.

\begin{figure}
\includegraphics[width=90mm]{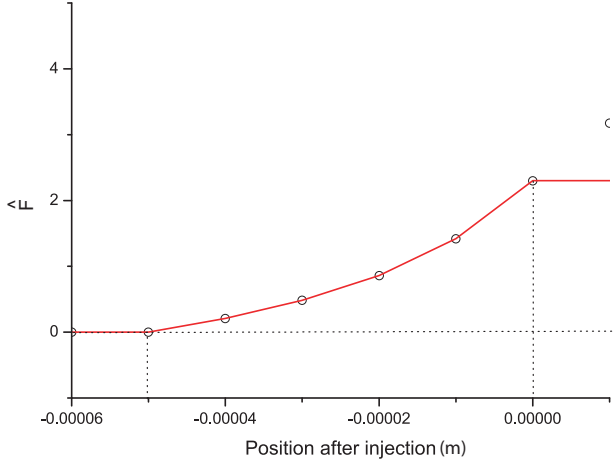}% Here is how to import EPS art
\caption{Normalized radial force $\hat{F}$ from the head of the
bunch ($\Delta s < 0$) to the tail as the bunch progresses inside
the bend ($z=0$ corresponds to the injection of the test particle
in the magnet). Here $R=1 m, \gamma = 100, h = 10 \mu m$ and the
bunch length is $200 \mu m$.The test particle is located at
$\Delta s = 0$.
%Curve A corresponds to the simulation by
%$TraFiC^4$, curve B is the contribution, given by Eq.
%\ref{HTbunch}, of the particles with $\Delta s < -h$; the curve
%A-B is the difference between the latter curve, giving the
%contribution from the particles with $\Delta s > -h$ .
\label{FIG6}}
\end{figure}
This reasoning can explain pretty well the features in Fig.
\ref{FIG6}. The radial force grows exactly as described in Eq.
(\ref{HTbunch}) starting from about $z=-l_b/4=50~\mu m$ until
$z=0$ (which is roughly the first formation length). Note that
with the source located at $\Delta s = h$ corresponds a retarded
angle $\phi=0$, which means that the electromagnetic signal from
this source, which is emitted at $z=0$, will reach the test
particle when this has also the position $z=0$. After $z=0$, the
test particle begins to interact with the sources located at
$\Delta s > -h$ while the ones at $\Delta s < -h$ reach a steady
state, in the sense that they keep on interacting in the same way
with the test electron.

\begin{figure}
\includegraphics[width=90mm]{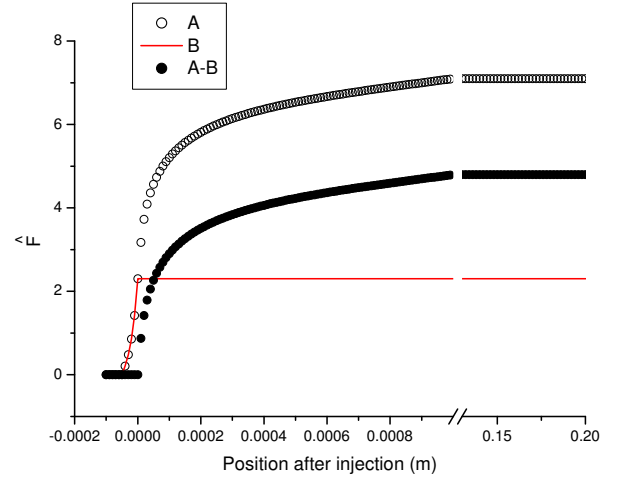}% Here is how to import EPS art
\caption{The same as Fig. \ref{FIG6} for a different range.
Normalized radial force from the head of the bunch ($\Delta s <
0$) to the tail as the bunch progresses inside the bend ($z=0$
corresponds to the injection of the test particle in the magnet).
Here $R=1 m, \gamma = 100, h = 10 \mu m$ and the bunch length is
$200 \mu m$. The test particle is located at $\Delta s = 0$. Curve
A corresponds to the simulation by $TraFiC^4$, curve B is the
contribution, given by Eq. \ref{HTbunch}, of the particles with
$\Delta s < -h$; A-B is the difference between the two curves,
ascribed to the contribution from the particles with $\Delta s
> -h$.\label{FIG7}}
\end{figure}

This is illustrated by Fig. \ref{FIG7}, where the normalized
radial force $\hat F$ is plotted.

The difference between the simulation results and the analytical
estimation in Eq. (\ref{HTbunch}) gives, quantitatively, the
radial interaction due to the sources at $\Delta s > -h$. For $h =
10 ~\mu m$ and $\gamma = 100$, as in our case, one expects a
second formation length equal to $10^{-3} ~m$ which is exactly
what one gets: the actual data show, in fact, that a maximum is
reached when $z = 0.001~ m$ at $\hat{F} \simeq 7.098$.

%\begin{figure}
%\includegraphics*[width=90mm]{HTlong.eps}% Here is how to import EPS art
%\caption{Normalized radial force from the head of the bunch
%($\Delta s < 0$) to the tail as the bunch progresses inside the
%bend ($z=0$ corresponds to the injection of the test particle in
%the magnet). Here $R=1 m, \gamma = 100, h = 10 \mu m$ and the
%bunch length is $200 \mu m$. The test particle is located at
%$\Delta s = 0$. Here we propose a comparison between the steady
%state contribution by $TraFiC^4$ (circles), which includes all the
%particles with $\Delta s < 0$ and the steady case from Eq.
%(\ref{HTbunch}) (solid line) which gives the contributions of the
%sources at $\Delta s < -h$. \label{FIG8}}
%\end{figure}
%

\begin{figure}
\includegraphics[width=90mm]{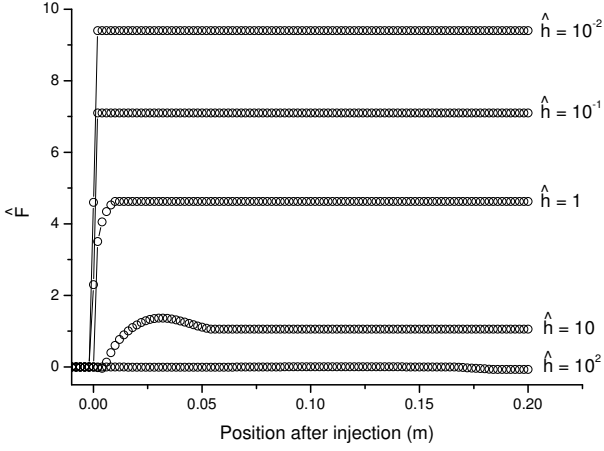}% Here is how to import EPS art
\caption{Normalized radial force from the head of the bunch
($\Delta s < 0$) to the tail as the bunch progresses inside the
bend ($z=0$ corresponds to the injection of the test particle in
the magnet). Here $R=1 m$ and $\gamma = 100$. Here we plot the
results from $TraFiC^4$ for several values of $\hat{h}$. The bunch
length is $200 \mu m$. The test particle is located at $\Delta s =
0$.\label{FIG9}}
\end{figure}
Fig. \ref{FIG7} also shows the steady state regime, when all the
particles in front of the test particle interact with the latter
from inside the bend. The solid line in Fig. \ref{FIG7} shows,
once again, the interaction with the sources located at $\Delta s
< -h$. Finally, in Fig. \ref{FIG9} we present the results from
$TraFiC^4$ for several values of $h$. Of course, when $h > l_b$,
i.e. $h > 100 ~\mu m$ there are no sources characterized by
$\Delta s < -h$ at all.

\begin{figure}[h]
\includegraphics[width=80mm]{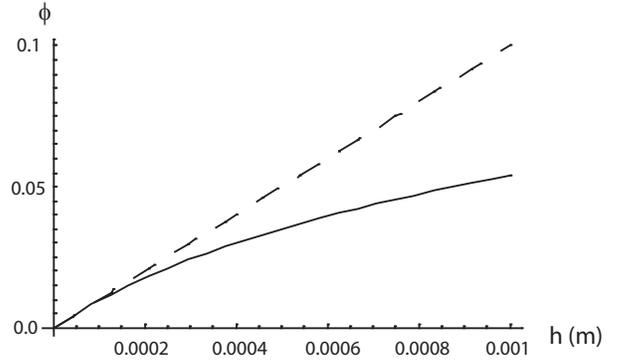}% Here is how to import EPS art
\caption{Solution of the retardation condition (solid line) at
$\Delta s = 0$ as a function of $h$ and comparison with $\phi =
\gamma h$ (dashed line) \label{FIG10}}
\end{figure}

One may expect that the system enters the steady state at $z
\simeq \gamma h$, which is correct for $h = 1 ~\mu m$, $h = 10
~\mu m$ and, by figure inspection, for $h = 100 ~\mu m$.
Nevertheless, in the case $h = 1 ~mm$ and $h = 10~ mm$ the system
enters the steady state at, respectively, $z \simeq 0.054~ m$ and
$z \simeq 0.186 ~m$, which are clearly smaller than $\gamma h$.
The reason for this apparent discrepancy is due to the
approximation $\sin(\phi/2) \simeq \phi/2 $ which has been used to
derive the rule of thumb $L_2 \simeq \gamma h$ starting from the
retardation condition Eq. (\ref{RETCONDH}). A comparison between
the rule of thumb proposed before (dashed line) and the real
solution of the retardation condition (solid line) is given in
Fig. \ref{FIG10}: one can easily see that, when $h = 1 ~mm$, $\phi
\simeq 0.054$. In the same way, at $h = 10 ~mm$, $\phi \simeq
0.186$ (remember that $R = 1 m$).

By comparison, in Fig. \ref{FIG6}, between the contributions to
the head-tail interaction in the cases $\Delta s < -h$ and
$-h<\Delta s<0$ the reader can see that these are comparable in
magnitude, although $h$ is much shorter than the bunch length. A
qualitative explanation of this fact can be given. When $\Delta s
\ll \phi R$, so that it can in fact be neglected in the
retardation condition, one gets $\phi \simeq \gamma h/R$. Now, the
force felt by the test particle in the zeroth order in $\phi$ is
proportional to $\gamma^3 /(R^2 \hat \phi )\simeq \gamma / (R h)$:
simple considerations suggest that this order of magnitude will
not change appreciably up to $\Delta s$ of order $-h/\gamma$. As a
result, the region $(0,-h/\gamma)$ gives a contribution, after
integration, proportional, approximately, to $\lambda/R$. On the
other hand, the maximum interaction in the region $\Delta s<-h$ is
of the order of $e /(Rh)$, which also corresponds, after
integration, to a magnitude proportional to $\lambda/R$. This
explains the reason why a small region $-h<\Delta s<0$ can give
contributions of the same order of magnitude as the much longer
interval in which $\Delta s < -h$.

\subsection{Tail-Head interaction}

We will now analyze the tail-head part of the interaction when
$\Delta s > 0$. In the case where the bunch enters the bend we
have contributions from retarded sources both in the bend and in
the straight line before the bend. The contribution from the
retarded sources in the magnet can be obtained by simple
integration of Eq. (\ref{Totexp}), and reads

\begin{widetext}
\begin{equation}
{F_\mathrm{\bot m}^\mathrm{B}} \simeq  { e^2 \lambda_0 \over {4
\pi \varepsilon_\mathrm{0} R}} \left[
\ln\left({\hat{\phi}_\mathrm{max}\over{\hat{\phi}_\mathrm{min}}}\right)
+ {4(2 \hat h^2+\hat\phi^2_\mathrm{max})\over{4\hat h^2+4
\hat{\phi}^2_\mathrm{max}+\hat{\phi}^4_\mathrm{max}}}- {4(2 \hat
h^2+\hat\phi^2_\mathrm{min})\over{4\hat h^2+4
\hat{\phi}^2_\mathrm{min}+\hat{\phi}^4_\mathrm{min}}}\right]~,
\label{shortBR2}
\end{equation}
\end{widetext}
where "m" is a reminder that the contributions treated by Eq.
(\ref{shortBR2}) are all from the "magnet". All that is left to do
now, is to investigate the values which $\hat{\phi}_\mathrm{min}$
and $\hat{\phi}_\mathrm{max}$ assume.

Let us first define with $\hat \eta$ the normalized angular
position of the test particle inside the bending magnet. Now
define with $\hat{\phi}^*$ the solution of the retardation
equation $\Delta \hat{s}_\mathrm{min} = \hat{\phi}^*/2 +
\hat{\phi}^{*3}/24-\hat h^2/(2 \hat \phi^*)$. If $\hat{\phi}^* <
\hat{\eta}$, the retarded position of the first source particle is
in the bending magnet, and $\hat{\phi}_\mathrm{min} =
\hat{\phi}^*$. On the other hand, when $\hat{\phi}^* > \hat{\phi}$
there are no contributions to the radial force from the bend.

Next, we define with $\hat{\phi}^{**}$ the solution of $\Delta
\hat{s}_\mathrm{max} = \hat{\phi}^{**}/2 +
\hat{\phi}^{**3}/24-\hat h^2/(2 \hat \phi^{**})$ (in our case
$\Delta \hat{s}_\mathrm{max}$ will be equal to one half of the
normalized bunch length). Supposing $\hat{\phi}^* < \hat{\eta}$,
if $\hat{\phi}^{**} < \hat{\eta}$ too, then all the particles
contribute from the bend, and $\hat{\phi}_\mathrm{max} =
\hat{\phi}^{**}$. On the other hand, when $\hat{\phi}^{**} >
\hat{\eta}$, we have a mixed situation, in which part of the
particles contribute from the bend and others from the straight
line before the magnet. In this case $\hat{\phi}_\mathrm{max} =
\hat{\phi}$.

\begin{figure}[!htb]
\includegraphics[width=90mm]{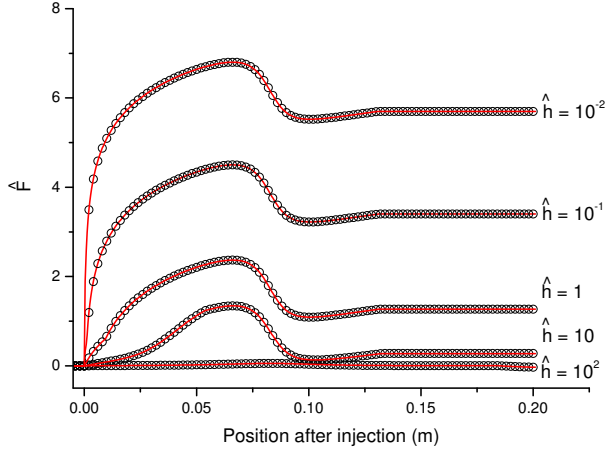}% Here is how to import EPS art
\caption{\label{FIG11} Normalized radial force acting on a test
particle from a bunch with rectangular density distribution
entering a hard-edge bending magnet as a function of the position
of the test particle inside the magnet. The solid lines show
analytical results; the circles describe the outcome from
$\mathrm{TRAFIC}^4$. We chose $\Delta s_\mathrm{max} = 100 \mu$m,
$\gamma = 100$, $R=1$ m; graphs are plotted for several values of
the parameter $\hat{h}$.}
\end{figure}
\begin{figure*}[!htb]
\includegraphics*[width=180mm]{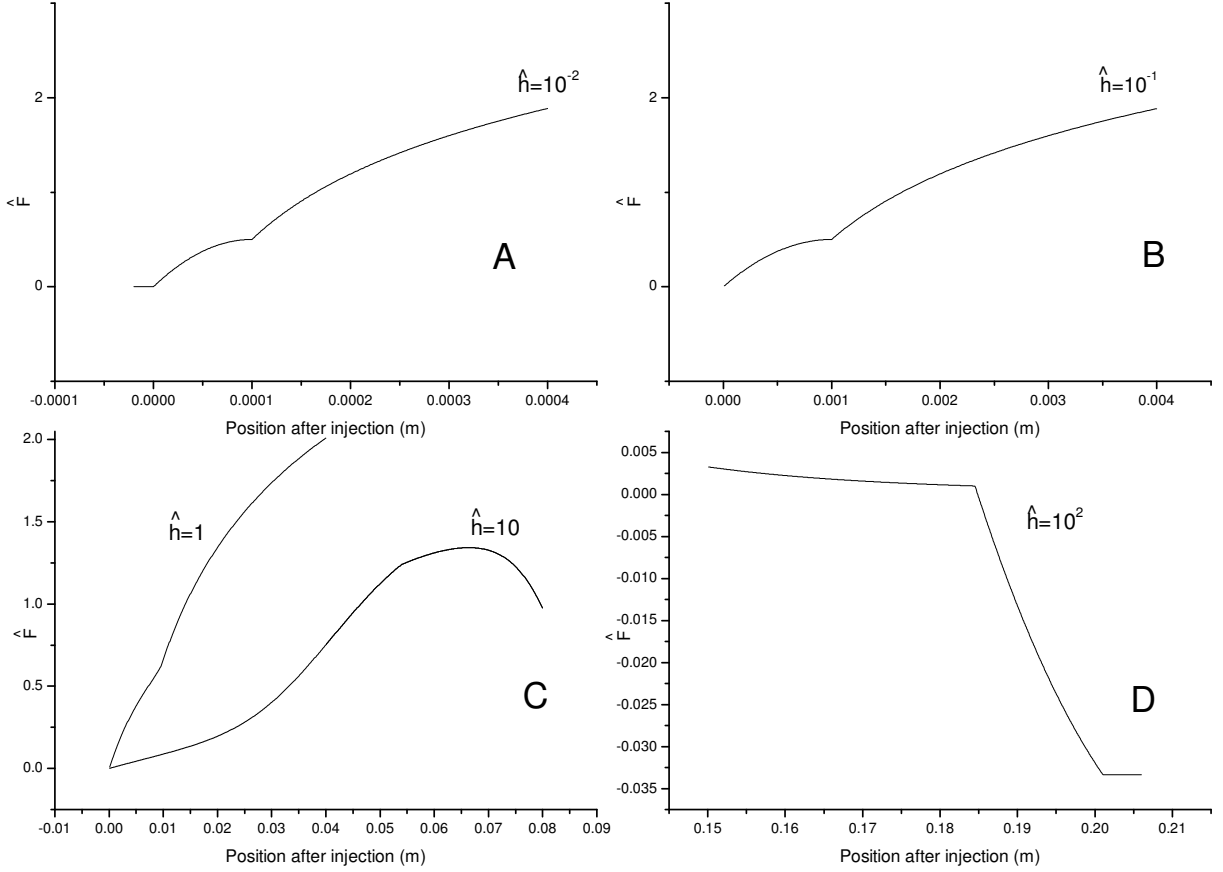}% Here is how to import EPS art
\caption{\label{FIG12} Normalized radial force (analytical result)
acting on a test particle from a bunch with rectangular density
distribution entering a hard-edge bending magnet as a function of
the position of the test particle inside the magnet. We chose
$\Delta s_\mathrm{max} = 100 \mu$m, $\gamma = 100$, $R=1$ m;
several curves for different values of $\hat{h}$ are shown. A
clear discontinuity in the first derivative is visible.}
\end{figure*}
The contribution from the retarded sources in the straight path
before the bend is given by

\begin{equation}
{F_\mathrm{\bot s}^\mathrm{B}} = \int_\mathrm{\Delta
\hat{s}_\mathrm{min}}^{\Delta \hat{s}_\mathrm{max}}
{R\over{\gamma^3}} F_\mathrm{\bot}(\hat{y}(\Delta \hat{s},\hat h),
\hat{\phi}) d\Delta \hat{s}~, \label{bunchcontrtrans}
\end{equation}
where "s" stands for "straight path", and where the expression for
$F_\mathrm{\bot}$ in the integrand is given by Eq. (\ref{expa}).
It is convenient, as done before, to switch the integration
variable from $\Delta \hat{s}$ to $\hat{y}$. The Jacobian of the
transformation is given by (see \cite{SALLONG})

\begin{equation}
{d\Delta \hat{s}\over{d \hat{y}}} \simeq
{(\hat{\phi}+\hat{y})^2+\hat{\phi}^4/4+\hat{h}^2\over{2
(\hat{\phi}+\hat{y})^2}} \label{Jacobiantransient}
\end{equation}
After substitution of Eq. (\ref{Jacobiantransient}) and Eq.
(\ref{expa}) in Eq. (\ref{bunchcontrtrans}), one can easily carry
out the integration, thus getting

\begin{eqnarray}
{F_\mathrm{\bot \mathrm{s}}^\mathrm{B}} \simeq  {2 e^2 \lambda_0
\over {4 \pi \varepsilon_\mathrm{0} R}} \Bigg[ {\hat{\phi}\left(4
\hat{y}_\mathrm{min} + 2 \hat{\phi} + \hat{\phi}^3 \right)\over{4
(\hat{y}_\mathrm{min}^2+\hat{h}^2) + 8
\hat{y}_\mathrm{min}\hat{\phi} + 4 \hat{\phi}^2 +
\hat{\phi}^4}}\nonumber\\&& \cr- {\hat{\phi}\left(4
\hat{y}_\mathrm{max} + 2 \hat{\phi} + \hat{\phi}^3 \right)\over{(4
\hat{y}_\mathrm{max}^2 + \hat{h}^2)+ 8
\hat{y}_\mathrm{max}\hat{\phi} + 4 \hat{\phi}^2 +
\hat{\phi}^4}}\Bigg]&&~. \label{bunchtransfinale}
\end{eqnarray}
As done before for $\hat{\phi}_\mathrm{min}$ and
$\hat{\phi}_\mathrm{max}$, we can now investigate the values of
$\hat{y}_\mathrm{min}$ and $\hat{y}_\mathrm{max}$.

Let us start with $\hat{y}_\mathrm{min}$. First, we define with
$\hat{y}^*$ the solution of the retardation condition $\Delta
\hat{s}_\mathrm{min} = (\hat{\phi}+\hat{y}^*)/2 +
(\hat{\phi}^{3}/24) (4 \hat{y}^*
+\hat{\phi})/({\hat{y}^*+\hat{\phi}})-\hat{h}^2/(2\hat{y}^*+2\hat
\phi)$. If $\hat{y}^*>0$, the retarded position of the first
source particle is in the straight line before the bending magnet,
and $\hat{y}_\mathrm{min} = \hat{y}^*$. On the other hand, when
$\hat{y}^* < 0$,  the retarded position of the first source
particle is in the bend, and $\hat{y}_\mathrm{min} = 0$.

Next, we define with $\hat{y}^{**}$ the solution of $\Delta
\hat{s}_\mathrm{max} = (\hat{\phi}+\hat{y}^{**})/2 +
(\hat{\phi}^{3}/24) (4 \hat{y}^{**}
+\hat{\phi})/({\hat{y}^{**}+\hat{\phi}}-\hat{h}^2/(2\hat{y}^{**}+2\hat
\phi))$ (again, in our case $\Delta \hat{s}_{max}$ is just half
the bunch length normalized to $R/\gamma^3$). Consider the case
$\hat{y}^{**} < 0$: all the particles contribute from the bend,
that is we entered the steady-state situation. In this case
$\hat{y}_\mathrm{max} = \hat{y}_\mathrm{min}=0$. On the other
hand, when $\hat{y}^{**} > 0$, we have again a mixed situation, in
which part of the particles contribute from the bend and others
from the straight line before the magnet. In this case
$\hat{y}_\mathrm{max} = \hat{y}^{**}$.

The following step is to actually plot the radial force exerted on
an electron by a bunch with rectangular distribution entering a
long bend.
%It is convenient to choose a different normalization
%factor from the earlier cases, and we therefore define $f = {e^2
%\lambda_0 /(4 \pi \varepsilon_\mathrm{0} R)} \ln(\Delta
%\hat{s}_\mathrm{max})$.
Our results, compared, once again, with simulations by
$\mathrm{TRAFIC}^4$, are shown in Fig. \ref{FIG11} for a bunch
length of $100~ \mu$m, $\gamma = 100$, $R=1$ m and for different
values of $\Delta \hat{s}$. A perfect agreement is obtained with
the results by $TraFiC^4$. A first characteristic length is
obviously given by the solution of the retardation condition with
$\Delta s = l_b/2$, and gives the position at which the test
particle begins to feel a steady interaction from the tail
sources. There is a second characteristic length more difficult to
see, though: a careful inspection of Fig. \ref{FIG11} shows, in
fact, a small irregularity (actually a discontinuity in the first
derivative) in the curve for $\hat{h} = 1$ at the position $z
\simeq 1~cm$. Such irregularities are present in all the curves in
Fig. \ref{FIG11}, although one has to look carefully for them by
magnifying parts of the plots, as shown in Figs. \ref{FIG12}A-D.
It is this fact which actually suggests the presence of a second
formation length.

When the value of $h$ is small the discontinuity is located,
approximately, at $z\simeq \gamma h$: for example when $h = 1~ \mu
m$, in Fig. \ref{FIG12}A, we have a discontinuity in the first
derivative of the curve at $z \simeq 0.1 ~mm$. Nevertheless, this
value changes as we increase h. In fact, when $h = 1~mm$, in Fig.
\ref{FIG12}C, the discontinuity is at $z \simeq 5.4 ~cm$, while
when $h = 1 ~cm$, in Fig. \ref{FIG12}D, we find a value of $z
\simeq 18.6 ~cm$: these are the same numerical values found when
discussing the entrance in the steady state of the Head-Tail
interaction in the previous subsection. The reader will remember
that these are, in fact, the solutions of Eq. (\ref{RETCONDH}).
From a physical viewpoint, the solution of Eq. (\ref{RETCONDH}) is
the position at which the test particle begins to feel the
electromagnetic signal from the source at $\Delta s = 0$.

Before that point, the test particle feels interaction from
particles behind it but only due to velocity fields, since the
retarded positions of all the electrons behind the test one are
not yet in the bend. After that particular point, the force on our
test electrons has a component due to the acceleration field too.
This suggests that the physical meaning of the presence of this
second formation length is simply the switching on of the
contribution of the acceleration field.  As a last remark it is
interesting to note that Tail-Head and Head-Tail interactions have
a characteristic length in common, but for completely different
reasons.

\section{\label{sec:concl}SUMMARY AND CONCLUSIONS}

In this paper we presented a fully electrodynamical study of
transverse self-forces within an electron bunch moving in an arc
of a circle in the case the test particle is endowed with a
vertical displacement $h$. We strived for a generalization of the
results obtained in \cite{OURS} in order to obtain a better
qualitative and quantitative explanation of the physics involved
in the problem and, in particular, to explain the behavior of the
self-interaction depicted in Fig. \ref{FIG1}.

First we generalized the results in \cite{OURS} in the case of a
two-particle system. Then, by integration of these results, the
case of a line bunch and a test particle with a vertical
displacement was studied. This case includes all the relevant
physics present in the situation of a bunch with vertical size.

Besides allowing one to generalize results obtained in
\cite{OURS}, our study aimed at a physical understanding of the
results by $TraFiC^4$: we found that the bunch can be divided into
four separate regions (over which one can integrate the
two-particle interaction) corresponding to four different types of
source-test interaction, namely head-tail with $|\Delta s| < h$,
head-tail with $|\Delta s|
> h$, tail-head with contributions from velocity fields alone, and
tail-head with contributions from both acceleration and velocity
field. We found that these regions correspond to four
characteristic formation lengths, which can be determined
quantitatively by simple analytical calculations.

For the first, the third, and the fourth region, we could use
relatively simple analytical results in order to describe the
situation and easily perform crosschecking with $TraFiC^4$. The
perfect agreement we found gives us much information: this
constitutes, in fact, a reliable cross-check which provides, at
the same time, very strong indication that the code calculates the
self-interaction in the proper way and that the approximations
which we made in our analytical theory are correct. Moreover, we
have physical explanations of the self-interactions in terms of
formation-lengths and type of source-test interaction.

Because of mathematical difficulties linked with the structure of
the retardation condition we left the quantitative discussion of
the second region ($-h < \Delta s < 0$) for future work.
Nevertheless we were able to understand the physical meaning of
this region and to determine its formation length.
\newpage

\section{\label{sec:acknowl}ACKNOWLEDGEMENTS}

The authors wish to thank Reinhard Brinkmann, Yaroslav Derbenev,
Klaus Floettmann, Vladimir Goloviznin, Georg Hoffstaetter, Rui Li,
Torsten Limberg, Jom Luiten, Helmut Mais, Philippe Piot, Joerg
Rossbach, Theo Schep and Frank Stulle for their interest in this
work.

\end{document}